\documentclass[onecollarge,natbib]{svjour2}
\bibpunct{[}{]}{;}{n}{}{,} % to get "[numbered]" references from natbib
\smartqed  % flush right qed marks, e.g. at end of proof
\usepackage{graphicx,color}
\usepackage[normalem]{ulem}
\newcommand{\cd}{\makebox[0.08cm]{$\cdot$}}

\journalname{Few-Body Systems}
\begin{document}

\title{Current conservation in electrodisintegration of a bound system in the Bethe-Salpeter approach
%about the article that should go on the front page should be
%placed here. General acknowledgments should be placed at the end of the article.}
}
%\subtitle{Do you have a subtitle?\\ If so, write it here}

%\titlerunning{Short form of title}        % if too long for running head

\author{V.A.~Karmanov         \and
        J.~Carbonell %etc.
}

%\authorrunning{Short form of author list} % if too long for running head

\institute{V.A.~Karmanov \at
              Lebedev Physical Institute, Leninsky Prospekt 53, 119991
Moscow, Russia \\
              %Tel.: +123-45-678910\\
              %Fax: +123-45-678910\\
              \email{karmanov@sci.lebedev.ru}           %  \\
%             \emph{Present address:} of F. Author  %  if needed
           \and
           J.~Carbonell \at
              Institut de Physique Nucl\'eaire,
Universit\'e Paris-Sud, IN2P3-CNRS, 91406 Orsay Cedex, France
}

\date{Received: date / Accepted: date}
% The correct dates will be entered by the editor

\maketitle

\begin{abstract}
Using our solutions of the Bethe-Salpeter equation with OBE kernel in Minkowski space both for the bound and scattering states, we calculate the transition form factors for electrodisintegration of the bound system which  determine the electromagnetic current $J$ of this process.
Special emphasis is put  on verifying the gauge invariance which should manifest itself in the current conservation. 
We find  that for any value of the momentum transfer $q$ the contributions of the plane wave and the final state interaction to the quantity $J\cdot q$ cancel each other thus providing $J\cdot q=0$. 
However, this cancellation
is obtained only if the initial Bethe-Salpeter amplitude (bound state), the final  one (scattering state) and the current operator are strictly consistent with each other. 
A reliable result for the transition form factor can be  found only in this case.
% \and More}
\end{abstract}

%%%%%%%%%%%%%%%%%%%%%%%%%%%%%%
\section{Introduction} \label{intro}
Bethe-Salpeter (BS) equation \cite{bs} provides an efficient theoretical framework to describe bound and scattering states of a relativistic system.
Finding its solution  is complicated by the singularities in the integrand of the equation as well as by the singular character of the amplitude itself. 
To avoid this difficulty, one can perform  Wick rotation \cite{WICK_54}
and transform the BS equation in the Euclidean form.
However, the Wick rotation cannot be performed in the integral for electromagnetic  (e.m.) form factors (see e.g. \cite{ck-trento}). 
Therefore, to calculate form factors, we need the BS solution in Minkowski space.  
In finding these solutions, an important progress was achieved  in the recent years  using different and independent methods (see \cite{LC2013-2} for a brief review). 

In one of these methods \cite{burov}  the kernel of the BS equation is approximately represented in a separable form. 
This allows to perform more advanced analytical calculations  and therefore simplifies finding solutions and form factors. 
Another method is based on representing  the BS amplitude via the Nakanishi integral  \cite{nak63}  both for  bound 
\cite{KusPRD95,KusPRD97,bs1,bs2,Sauli_JPG08} and  scattering \cite{fsv-2012}  states.
The elastic e.m. form factor in this method was calculated in \cite{ckm_ejpa}.    

Recently we developed a method  \cite{CK_PLB_PRD}  based on the direct treatment of singularities of the BS equation. Both the bound
and scattering  state  amplitudes in Minkowski space were found. They allow to calculate \cite{transit} the electrodisintegration of the bound system, i.e.,   form factor of the transition: bound  $\to$ scattering state. The contribution of the final state interaction (FSI) to this form factor is given by the Feynman graph shown in fig.  \ref{triangle} (left panel). The left and right vertices in this graph are just 
the Minkowski BS amplitudes for the bound (left) and scattering (right) states.  

In this paper we will analyze the conservation of the calculated e.m. current in the inelastic transition. We will see that it is indeed conserved
(as it should be), however this conservation is due to rather delicate cancellation between the plane wave (PW) contribution (right panel of fig. \ref{triangle}) and FSI  (left panel of fig. \ref{triangle})
which requires consistency between the bound and scattering state solutions and also the e.m. current operator. 
All of them  should be found in the same dynamical framework. 
This provides a strong test of all these quantities simultaneously and, especially, an important test of the transition form factor. 

The need for an internal consistency required to ensure the gauge invariance was extensively  discussed in \cite{Adam_V_Orden_Gross} in the framework of the BS and the Gross spectator equations. Our numerical results are in agreement with this general expectation. If this consistency and, hence, the gauge invariance are violated, the consequence of that is not reduced only to appearance in the current of a non-conserved part (which, however,  drops out from the cross section at all). We emphasize that the transition form factors, extracted from the conserved part of the non-conserved current, are also deficient.

We consider  in this work spinless particles. The generalization to the fermion case is straightforward and does not contain new problems, since the fermionic and spineless propagators have the same singularities.

%%%%%%%%%%%%%%%%%%%%%%%%%%%%%%%%
\section{Transition form factors}\label{tff}

For spinless particles, the general decomposition of the transition current $J_{\mu}$ in terms of form factors, non-assuming current conservation, has the form:
\begin{equation}\label{ffc}
J_{\mu}=\left[(p_{\mu}+p'_{\mu})+ (p'_{\mu}-p_{\mu})\frac{Q_c^2}{Q^2}\right]F(Q^2)- (p'_{\mu}-p_{\mu})\frac{Q_c^2}{Q^2}F'(Q^2),
\end{equation}
where $Q^2=-(p-p')^2$ and $Q_c^2=  {M'}^2-M^2$. $M$ is the bound state mass, $M'$ is the invariant mass of the final (scattering) state.
The factor $\frac{Q_c^2}{Q^2}$ in the last term is introduced for convenience, to ensure  some scale properties to the form factor $F'(Q^2)$. The decomposition (\ref{ffc}) implies that the initial and final states have total zero angular momenta, i.e. they are composed from the S-waves only.

The first term in (\ref{ffc}) satisfies the current conservation, whereas the second one  does not, that is:
\begin{equation}\label{cconserv}
q\cdot J= Q_c^2F'(Q^2)\neq 0,\quad \mbox{if $F'(Q^2)\neq 0$}. 
\end{equation} 
In the elastic case the last term in (\ref{ffc}) is forbidden by the symmetry between initial and final states. In the inelastic case, this term should be zero due to the current conservation. 

Below we will calculate both form factors $F(Q^2)$ and $F'(Q^2)$. Note that the sum of the two graphs shown in fig. \ref{triangle} exhausts all the contributions to the e.m. current  in the model with OBE kernel, 
i.e., it provides the full e.m. current.
We will check by  numerical calculations whether this full current  is conserved or not, i.e., whether $F'(Q^2)=0$. 
We will see  that the current conservation is a rather subtle property which is violated by approximations in the solutions or by inconsistencies in the input quantities,
 like e.g. initial and final BS amplitudes and/or e.m. current operator.  

 %%%%%%%%%%%%%%%%%%%%%%%%%%%%%%%%%%
\section{Calculating form factors}\label{fsi}

\begin{figure}[hbtp] 
\centering
\includegraphics[width=6.5cm]{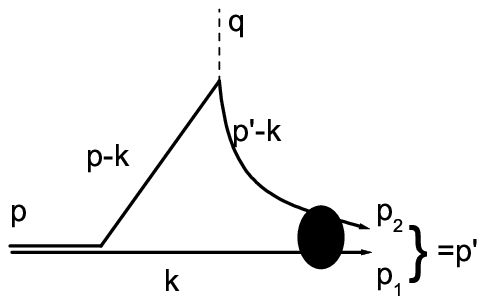} \hspace{0.8cm}
\includegraphics[width=6.5cm]{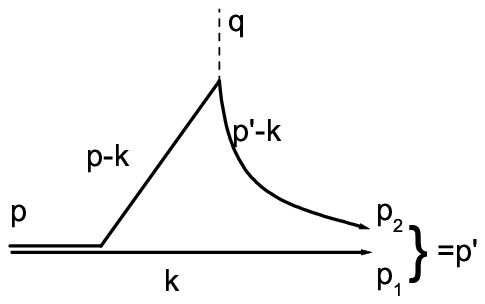}
\caption{Feynman graph FSI (left panel) and PW contribution (right panel) for the transition e.m. form factor.}\label{triangle}
\end{figure}

It is convenient to carry out calculations in the reference system  where $p'_0=p_0$ (i.e., $q_0=0$) and $\vec{p}$ and $\vec{p'}$ are collinear (i.e., they are either parallel or anti-parallel to each other, depending on the value of the momentum transfer $Q^2$). 
In the elastic case this system coincides with the Breit frame
$\vec{p}+\vec{p'}=0$ (and so $|\vec{p}|=|\vec{p'}|$,  $p'_0=p_0$). 
In inelastic case and in the frame with $p'_0=p_0$ we have $|\vec{p}|\neq|\vec{p'}|$.  
If in this frame we take the zero component of the current to find the form factor $F(Q^2)$
\begin{equation}\label{JF}
J_{0}=2p_{0}F(Q^2).
\end{equation}

We first calculate the FSI contribution.  The current $J_{\mu}$
is obtained by applying the Feynman rules to the graph in left panel of fig.   \ref{triangle}.
We follow the Feynman rules in the convention of Itzykson-Zuber \cite{IZ}. 
Taking the component $J_{0}$ from eq. (\ref{JF}) we find:
\begin{equation}\label{ffin1}
F_{FSI}(Q^2)=i\int \frac{d^4k}{(2\pi)^4}\,
\frac{(p_0-k_0)}{p_0}\frac{\Gamma_i \left(\frac{1}{2}p -k,p\right)\Gamma_f
\left(\frac{1}{2}p'-k,p'\right)} {(k^2-m^2+i\epsilon)[(p-k)^2-m^2+i\epsilon][(p'-k)^2-m^2+i\epsilon]}\, .
\end{equation}
The factor $(p_0-k_0)/p_0$ comes from the zero component of the four-vector $p_{\mu}+p'_{\mu}-2k_{\mu}$, corresponding to the upper (e.m.) vertex of the FSI graph in fig. \ref{triangle} 
under the condition $q_0=0$ and after dividing this component by  $2p_{0}$ from (\ref{JF}).
$\Gamma_i$ is the initial (bound state) vertex, $\Gamma_f$ is the final vertex (half-off-shell scattering BS amplitude). As mentioned, both S-wave vertices were found numerically in \cite{CK_PLB_PRD}. 
More precisely, in the assumed normalization convention, the function $\Gamma_f$  is related to the  S-wave  scattering solution  $F_0$ calculated in  \cite{CK_PLB_PRD} by $\Gamma_f=16\pi F_0$.

If the integral   (\ref{ffin1}) would not contain $\Gamma_i,\Gamma_f$, it can be calculated by applying to the product  of propagators the Feynman parametrization and then the Wick rotation. 
However, since the initial and final BS amplitudes $\Gamma_i,\Gamma_f$ in (\ref{ffin1}) are  known numerically,
the  Feynman parametrization cannot be applied to the full integrand in  (\ref{ffin1}) and  this 4D singular integral must be computed numerically as well. 
This calculation requires some  treatment of the pole singularities of propagators. For this aim, each pole contribution (two for each propagator, six  in total)  is represented as sum of a principal value and a delta-function. 
The integrals containing the delta-functions are evaluated analytically.  For the terms having the pole singularities (and also other, weaker singularities in the vertex functions  $\Gamma_i ,$ $\Gamma_f$) we 
use the subtraction technique. The terms, where the pole singularities are canceled by subtractions, are integrated numerically, 
whereas  the integrals from the added (pole) terms are calculated analytically. 
After that, the pole singularities in the added terms turn into the logarithmic ones. The latter ones can be safely integrated using appropiate numerical methods (as we did in \cite{CK_PLB_PRD}) or by  "brut force".  
The details of this calculation will be published in a future paper. 

The PW contribution to the current $J_{\mu}$ is obtained by applying the Feynman rules to the right panel of fig. \ref{triangle}. 
Like in the case of FSI,  the corresponding form factor is extracted from the component $J_0$ by eq.   (\ref{JF}) in the reference system  where $q_0=0$. That is:
\begin{equation}\label{Fpw}
F_{PW}=-\int d^4k \,\frac{(p_0-k_0)}{p_0}\frac{\Gamma_i\left(\frac{p}{2}-k,p\right)}{[(p-k)^2-m^2+i\epsilon]}\, \int\delta^{(4)}\left(k-p_s-\frac{p'}{2}\right)\frac{d\Omega_{\vec{p}_s}}{4\pi} \, .
\end{equation}
We keep the integral over $d^4k$ and the delta-function for the four-momenta conservation. This delta-function   replaces now the final BS amplitude $\Gamma_f$. 
When calculating the FSI contribution,  we decomposed the final state BS amplitude in partial waves and took into account the S-wave only. 
This is equivalent to averaging $\Gamma_f$ over the solid angle of the vector $\vec{p_s}$ in the rest frame $\vec{p'}=0$. 
 To extract the S-wave from the final plane wave we introduced in (\ref{Fpw}) the integration   over $\frac{d\Omega_{\vec{p}_s}}{4\pi}$.

The total form factor $F(Q^2)$ is then obtained as a sum of  the two contributions (\ref{ffin1}) and  (\ref{Fpw}):
\[F(Q^2)=F_{FSI}(Q^2)+F_{PW}(Q^2).\]

Now we will find the form factor $F'(Q^2)$.
According to eq. (\ref{ffc}), this form factor  is extracted from the e.m. current by:
\begin{equation}\label{Fp}
F'(Q^2)=\frac{J\cdot q}{Q_c^2}\, .
\end{equation}
As mentioned above the expression  for $J_{\mu}$ contains the four-vector  $(p_{\mu}+p'_{\mu}-2k_{\mu})$. After multiplying it by $q_{\mu}/Q_c^2$, we get the factor
$$
\frac{1}{Q_c^2}(p'-p)\cd(p+p'-2k)=\left.\frac{1}{Q_c^2}[{M'}^2-M^2-2(p'-p)\cd k]\right|_{p'_0=p_0}=\left(1-2\frac{\sqrt{Q^2}zk}{Q_c^2}\right),
$$
where $z$  is cosine of the angle between the integration variable $\vec{k}$ and the momentum $\vec{p}$ of the initial (bound state) system in the reference frame where $q_0=0$.
This factor appears instead of $(p_0-k_0)/p_0$ in (\ref{ffin1}) and  (\ref{Fpw}). Therefore,  two contributions $F'_{FSI}(Q^2)$ and $F'_{PW}(Q^2)$ in
the form factor  $F'(Q^2)=F_{FSI}'(Q^2)+F_{PW}'(Q^2)$
are determined by eqs. (\ref{ffin1}) and  (\ref{Fpw}) with  the replacement 
\begin{equation}\label{factor1}
\frac{(p_0-k_0)}{p_0} \to \left(1-2\frac{\sqrt{Q^2}zk}{Q_c^2}\right).
\end{equation}

\begin{figure}[htbp]
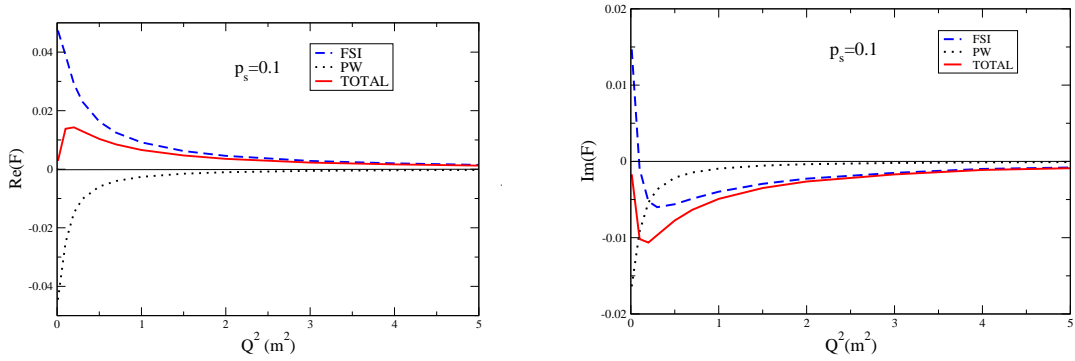

\centering
\includegraphics[width=6.5cm]{fig2a.eps} \hspace{0.8cm}
\includegraphics[width=6.5cm]{fig2b.eps}
\caption{(color online). Transition e.m. form factor $F(Q^2)$ as a function of $Q^2$. Initial (bound) state corresponds to the binding energy $B=0.01\,m$; final (scattering) state corresponds to the relative momentum $p_s=0.1\,m$ ($M'=2.00998$). FSI contribution is shown by the dashed curve, PW one by dotted curve and total form factor  by  solid curve. Left panel is the real part of form factor. Right panel is the imaginary part.}\label{ff}
\end{figure}
\begin{figure}[!h]
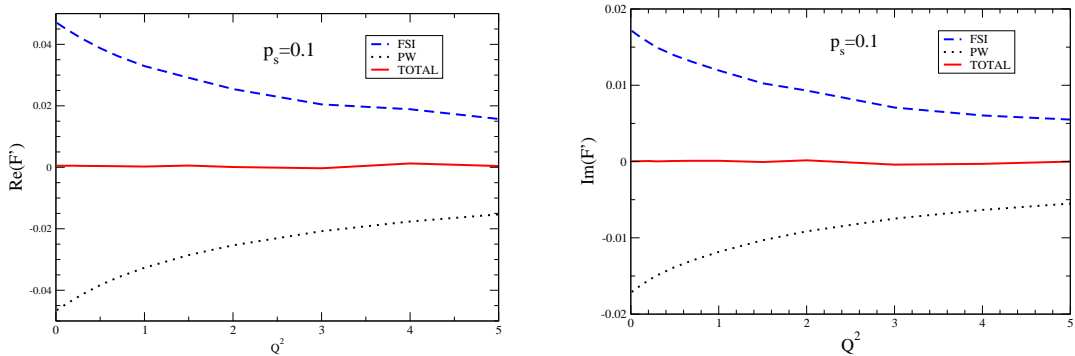

\centering
\includegraphics[width=6.5cm]{fig3a.eps} \hspace{0.8cm}
\includegraphics[width=6.5cm]{fig3b.eps}
\caption{(color online). Transition e.m. form factor $F'(Q^2)$ as a function of $Q^2$  for $p_s=0.1$. Parameters and notations are the same as in fig. \protect{\ref{ff}}.}
\label{figFp}
\end{figure}

%%%%%%%%%%%%%%%%%%%%%%%
\section{Numerical results}

Contrary to the elastic scattering, the inelastic (transition) form factor is complex. Its real and imaginary parts  as  function of $Q^2$ are shown in fig. \ref{ff}  at the left and right panels correspondingly.    
The  initial (bound) state corresponds to a binding energy $B=0.01\,m$ ($M=1.99\,m$), the final (scattering) state corresponds to a relative momentum $p_s=0.1\,m$ ($M'=2\sqrt{p_s^2+m^2}=2.00998\,m$).
The coupling constant in the OBE kernel is $\alpha=1.437$.
One can see that at relatively small momentum transfer $Q^2 < 1$ both contributions -- FSI and PW -- are important and they considerably cancel each other.
Whereas, at high momentum transfer the re-scattering (i.e., FSI) determines the tail of form factor and dominates over PW. Calculations for larger relative momentum $p_s=0.5\,m$ (the final state mass: $M'=2.336$) do not indicate on the predominance of FSI at large momentum transfer.

As mentioned above, the form factor $F'(Q^2)$ (equal to zero if the current is conserved) is obtained from $F(Q^2)$ by the replacement (\ref{factor1}) in the integrands (\ref{ffin1}) and  (\ref{Fpw}). Calculated in this way, the form factor $F'(Q^2)$  for $p_s=0.1$ 
is shown in fig. \ref{figFp}.
We see that the value of $F'(Q^2)$, in comparison to $F(Q^2)$ (fig. \ref{ff}), is practically indistinguishable from zero. This zero value is  however 
obtained as cancellation of FSI and PW contributions. The accuracy of our calculations provides  that the sum of two contributions $F'_{FSI}(Q^2)+F'_{PW}(Q^2)$ is by two-three 
orders of magnitude smaller than each contribution. We conclude that in the considered model the current is conserved with the precision of the numerical solutions. 
\bigskip

\begin{figure}[htbp]
\centering
\includegraphics[width=6.5cm]{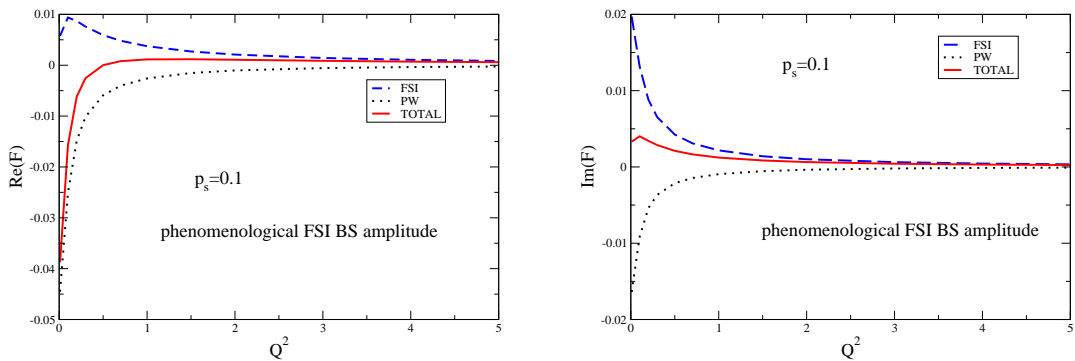} \hspace{0.8cm}
\includegraphics[width=6.5cm]{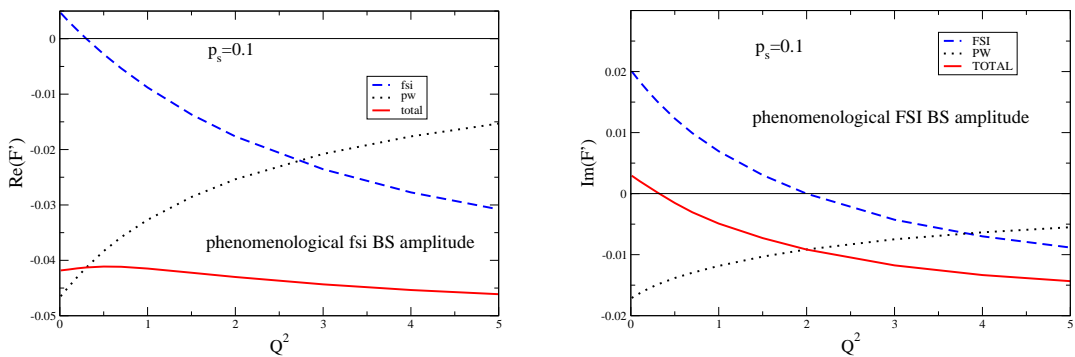}
\caption{(color online). Transition e.m. form factor $F(Q^2)$ as a function of $Q^2$  for $p_s=0.1$ calculated with the "phenomenological" FSI function 
(\ref{G_phenom}).  The notations are the same as in fig. \protect{\ref{ff}}.}
\label{F_phenom}
\end{figure}
\vspace{-0.3cm}
\begin{figure}[htbp]
\centering
\includegraphics[width=6.5cm]{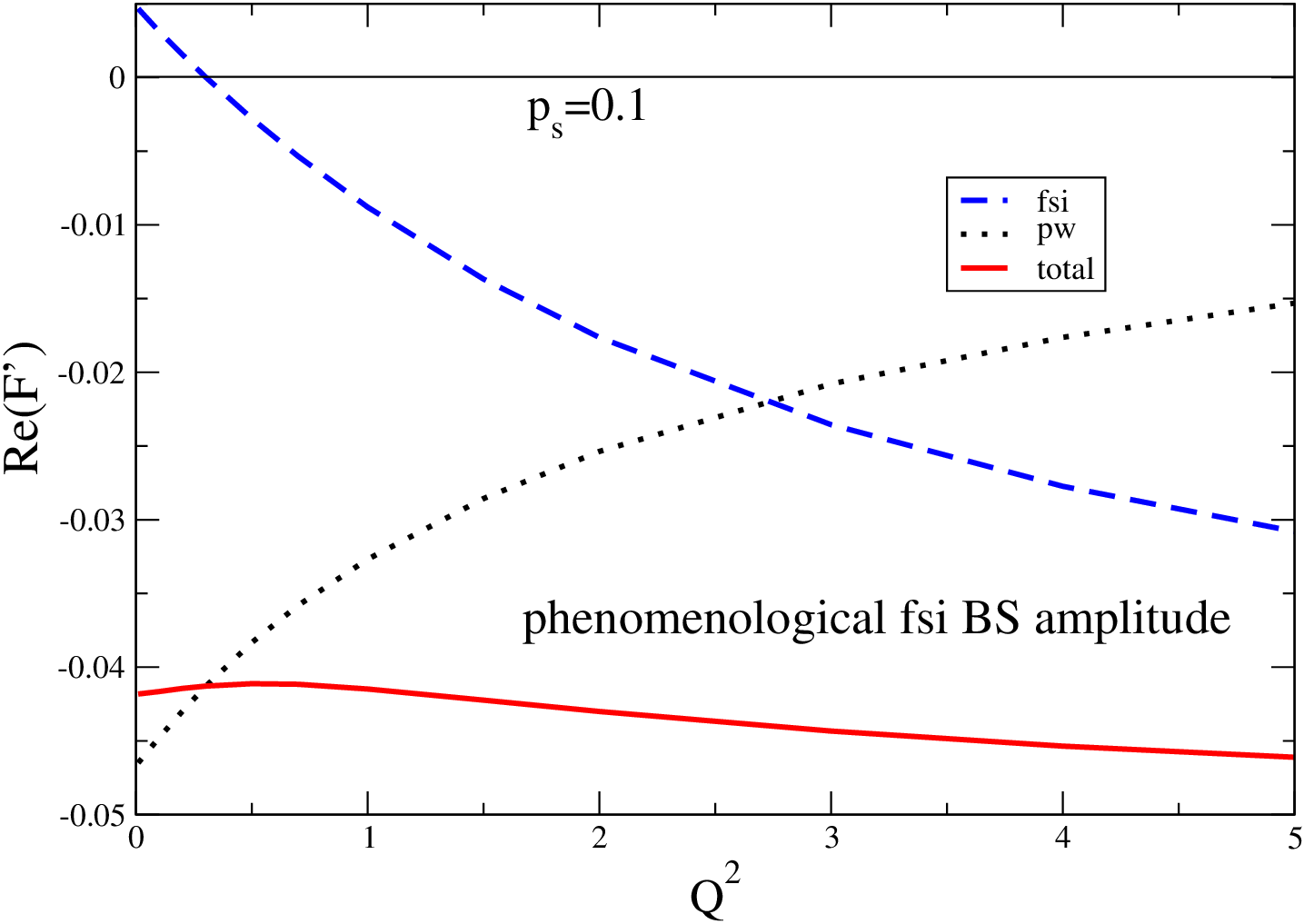} \hspace{0.8cm}
\includegraphics[width=6.5cm]{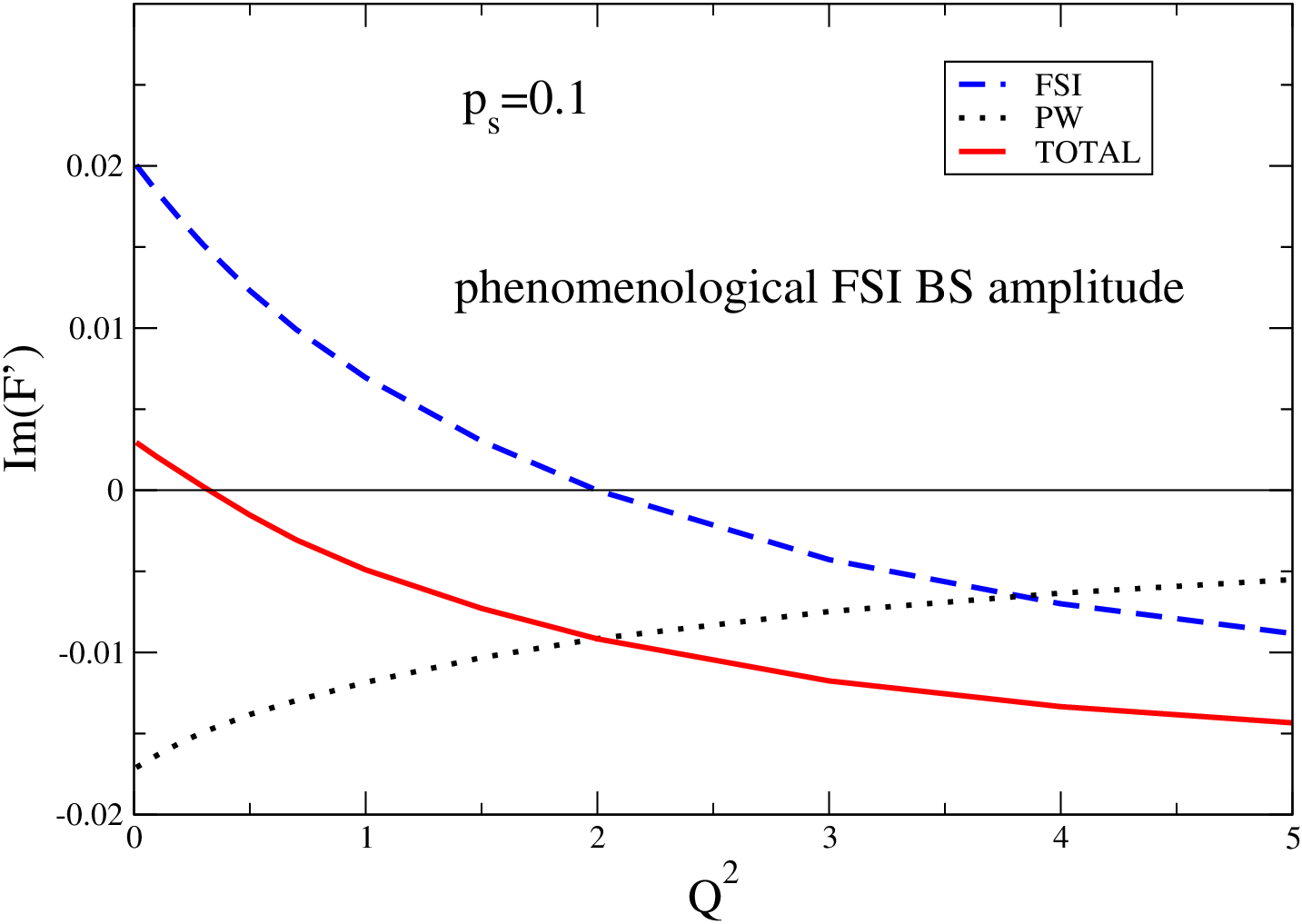}
\caption{(color online). Transition e.m. form factor $F'(Q^2)$ as a function of $Q^2$  for $p_s=0.1$ calculated with the "phenomenological" FSI function 
(\ref{G_phenom}).  The notations are the same as in fig. \protect{\ref{ff}}.}
\label{Fp_phenom}
\end{figure}

Though the current conservation is required from the first principles, 
the cancellation of FSI and PW contributions is a rather delicate point and therefore  can serve as a strong test of the quantities involved. 
Indeed,  both FSI and PW contributions contain the same initial bound state BS amplitude. At the same time, FSI contribution contains the final state BS amplitude, 
whereas the PW contribution -- does not. The cancellation of FSI and PW takes place provided both BS amplitudes and the current operator are consistent with each other, i.e. if they are correctly found 
in the same dynamical framework. To demonstrate that violation of this consistency indeed violates the current conservation, we replaced the final BS amplitude, found for the OBE kernel, by an "ad-hoc" function 
\begin{equation}\label{G_phenom}
G_f(k_0,k)\sim \frac{1}{(k_0^2+a^2)(k^2+b^2)}
\end{equation}
without changing the initial BS amplitude and the current. The form factor $F(Q^2)$ calculated with the function (\ref{G_phenom}) is shown in fig. \ref{F_phenom}. It has a typical behavior which, in itself, does not arouse any concern.

The transition form factor $F'(Q^2)$ calculated with the function (\ref{G_phenom})  for the parameters $a^2=1.5$, $b^2=1$ is shown in fig. \ref{Fp_phenom}. We see that it is not zero.
This means that the e.m. current  calculated with the "phenomenological" FSI function  (\ref{G_phenom}) is not correct. Therefore the form factor $F(Q^2)$,
fig. \ref{F_phenom}, extracted from this current, is also incorrect. 

%%%%%%%%%%%%%%%%%%%%% 
\section{Conclusion} 

Using the OBE model, we have found that if the bound and scattering state BS amplitudes as well as the  operator of e.m. current are exact solutions of the dynamical equations and consistent
with each other, the transition e.m. current is conserved. 
If this consistency is destroyed, the conservation is violated.  This violation results in two consequences. ({\it i}) The decomposition of the e.m. current in form factors obtains and additional contribution (the second term in eq. (\ref{ffc}) with the non-zero form factor $F'(Q^2)$) which does not satisfy the equality $J\cdot q=0$. 
However, the appearance of this term itself does not make any influence on observables, since $F'(Q^2)$ does not contribute in the cross section, due to conservation of the e.m. current of the incident electron. 
({\it ii}) The most important consequence is the fact that the non-conservation of the calculated e.m. current means its deficiency at all which prevents us from extracting a reliable transition form factor $F(Q^2)$. 
It is therefore mandatory, in practical calculations, like, for instance, in the deuteron electrodisintegration, to fulfill the current conservation.
If the form factor $F'(Q^2)$ responsible for non-conserved e.m. current is comparable to the physical ones ($F(Q^2)$ in the spineless case),   one can hardly trust to the calculated physical form factors too. 

The popular trick, consisting in the replacement of  the non-conserved current $J_{\mu}$  
by the conserved combination $J_{\mu}\to \tilde{J}_{\mu}= J_{\mu}-q_{\mu}(J\cdot q)/q^2$ does not solve the problem but only hides it. This new current $\tilde{J}_{\mu}$ satisfies the  current conservation for any arbitrary $J_{\mu}$, not only  for the correct one. If $J\cdot q\neq 0$, this indicates that the current  $J_{\mu}$ is defective at all. If so,  one can extract the correct transition form factor neither from $J_{\mu}$, nor from $\tilde{J}_{\mu}$. 

%%%%%%%%%%%%%%%%%%%%%%%%%%%%%%%%%%%%%%%%%%%%%%%%%%%%%%%

\end{document}